\documentstyle[11pt,aasms4]{article}
\lefthead{Cremonese et al.}
\righthead{Sodium from Hale-Bopp}

\begin{document}

\noindent
\title{Neutral sodium from comet Hale--Bopp: a third type of tail}

\author{G. Cremonese}
\affil{Osservatorio Astronomico, Vic. Osservatorio 5, 35122 Padova -- 
Italy}
\author{H. Boehnhardt\altaffilmark{1}}
\affil{University of Munich, Scheinerstr. 1, D--81679 Munich -- 
Germany}
\author{J. Crovisier and H. Rauer\altaffilmark{2}}
\affil{Observatoire de Paris--Meudon, 5 Place Jules Janssen, F--92195 
Meudon -- France}
\author{A. Fitzsimmons}
\affil{Queen's University, Belfast, BT7 1NN -- United Kingdom}
\author{M. Fulle}
\affil{Osservatorio Astronomico, via Tiepolo 11, I--34131 Trieste-- 
Italy}
\author{J. Licandro\altaffilmark{3}}
\affil{Instituto de Astrofis\`{i}ca de Canarias,
Via Lactea s/n, 38200 La Laguna -- Spain}
\author{D. Pollacco}
\affil{Isaac Newton Group, Apartado de Correos 368, Santa Cruz de La Palma, 38780 
Tenerife -- Spain}
\author{G.P. Tozzi}
\affil{Osservatorio Astrofisico di Arcetri, Largo E.Fermi 5, 50125 Firenze -- 
Italy}
\and
\author{R.M. West}
\affil{ESO, Karl--Schwarzschild--Str. 2, D--85748 Garching -- Germany}

\altaffiltext{1}{ESO, Santiago del Cile}
\altaffiltext{2}{Deutsche Forschungsanstalt f\"{u}r Luft und Raumfahrt,
Institut f\"{u}r Planetenerkundung, D--12484 Berlin -- Germany}
\altaffiltext{3}{Dept. de Astronom\`{\i}a, Universidad de la Rep\`ublica, 
Trist\`an Narvaja 1674, 11200 Montevideo -- Uruguay}

\begin{abstract}

We report on the discovery and analysis of a striking neutral sodium gas
tail associated with comet C/1995 O1 Hale-Bopp.
Sodium D--line emission has been observed at heliocentric distance
$r \leq 1.4$ AU in some long--period comets and the presence
of neutral sodium in the tailward direction
of a few bright comets has been noted, but the extent, and in particular
the source, has never been clear. 
Here we describe the first observations and analysis of a neutral
sodium gas tail in comet Hale--Bopp, entirely different from the
previously known ion and dust tails.  We show that the observed
characteristics of this third type of tail are consistent with it
being produced by radiation pressure due to resonance
fluorescence of sodium atoms and that the lifetime for
photoionization is consistent with recent theoretical calculations.

\end{abstract}


\section{Introduction}

\noindent
Sodium emission has been previously observed in both long-period and
dynamically new comets (\markcite{s16}Rahe et al. 1976, 
\markcite{s18}Sivaraman et al. 1979, \markcite{s6}Hicks 
and Fink 1997).
 Although the abundance of sodium in comets is much lower than other
observable  elements and molecules, it can be used as a tracer of
processes acting in the same environment. This is because the
extremely high efficiency of the sodium atom in resonant scattering of
solar radiation makes it detectable even when the column density is low.
For instance the absolute sodium abundance in the Io atmosphere is
$\sim$1\%, yet it has been the neutral species most studied
after its discovery and is an important tracer of the 
atmospheric interactions with the plasma torus (\markcite{s19}Thomas 1992).
Further examples are that of the tenuous atmospheres
of Mercury and the Moon. 
If the sources of the sodium observed in comet Hale-Bopp and in other 
comets can be identified, it could be possible to study the sodium 
distribution and dynamics as a tracer of physical processes that may
otherwise not be easily detectable.\\
\noindent
In the context of the present observations, the first possible
identification of a sodium tail was recorded by 
\markcite{s12}Newall (1910) during the spectroscopic
observation of comet C/1910 A1, made 
with a direct-vision prism inserted between the eye and the eyepiece of the 
25-inch telescope. Later a 7 degree-long sodium tail was reported
in comet C/1957 P1 Mrkos by \markcite{s13}Nguyen-Huu-Doan (1960)
using a Schmidt camera with an objective prism.
In both cases the sodium tail was identified
through spectroscopic observation of the D-line emission, however
there was no detailed analysis of the morphology of the tail and the
mechanism responsible for these features. More recently,
sodium was observed in comets C/1965f Ikeya-Seki (\markcite{sp}Spinrad and Miner, 1968, 
\markcite{h2}Huebner, 1970), C/1969 Y1 Bennett 
(\markcite{s16}Rahe et al. 1976) and C/1976 V1 West 
(\markcite{s17}Sivaraman et al. 1979) within 100000 km of the nucleus.
The Giotto spacecraft performed  {\it in--situ} measurements in
the coma of comet Halley and found the ratio Na/O 
$= 4.9\times10^{-3}$, close to the cosmic
ratio (\markcite{s9}Jessberger and Kissel, 1991). In this letter we report
on the discovery and analysis of an extensive  neutral sodium tail
observed in comet C/1995 O1 Hale-Bopp, first reported by
\markcite{s2}Cremonese (1997).\\

\section{Observations and data reduction}

\noindent
Images of Comet Hale--Bopp were obtained in the period 16--22 of April  
with the CoCAM wide--field imaging instrument on La Palma
plus a narrow--band sodium filter ($\lambda_c = 5892$\AA , FWHM$=15$\AA) 
designed to isolate emission from the sodium D$_2$ and D$_1$ lines.
CoCAM consists of a 35mm camera zoom lens working at f/3.5 and imaging
onto a $2220\times1180$ pixel EEV CCD chip, whose pixel size of
22.5$\mu$m square corresponds to 26 arcsec, thereby achieving on the
sky a total field of $17\times9$ degrees. Imaging was also
obtained with filters designed to isolate both H$_2$O$^+$ emission in
the ion tail ($\lambda_c = 6185$\AA , FWHM$=42$\AA ) and the solar
continuum due to dust scattering ($\lambda_c = 6250$\AA , FWHM$=25$\AA). 
The wide field sodium images immediately revealed the presence of a very
linear structure stretching several degrees close to the anti--sunward
direction, with no obvious counterpart in the H$_2$O$^+$ and dust images,
and which we took to be a separate neutral Na tail 
(\markcite{s2}Cremonese et al. 1997,
\markcite{s7}Fitzsimmons et al. 1997). An example image is shown in
figure 1.
\noindent
High--resolution spectroscopy was performed at several points along
this tail on April 19.9 UT, 20.9 UT and 23.9 UT,
using the 4.2m William Hershel Telescope with the Utrecht
Echelle Spectrograph and $1024\times1024$ pixel TEK CCD chip,
providing a dispersion of 0.053\AA\ pixel$^{-1}$ and a resolution
of $0.1$\AA\ with a slit width of $1.1$ arcsec. On the last 
date the narrow--band sodium filter described above was placed in the
spectrograph, thereby allowing long--slit spectroscopy to be performed. 
This configuration allowed us to have a larger spatial coverage of
$2.7$ arcmin while simultaneously isolating the single order with only
the NaD emission lines. In all spectra redshifted Na emission was
observed, thereby confirming the presence of neutral sodium in this tail.
Unfortunately it was not possible to obtain both images of the sodium tail
and high resolution spectra in the long slit configuration during the same
night, due to only one sodium interference filter being available.
The comet during  these nights had a heliocentric distance of 0.98 AU
and a geocentric distance of 1.61 AU.\\
\noindent
All the spectra were cosmic-ray cleaned and bias-subtracted using
FIGARO spectroscopic reduction routines 
(\markcite{s17}Shortridge et al. 1997).
For the normal echelle-mode data obtained on the 19th and 20th April, each
order was individually flatfielded using tungsten lamp
exposures at the same instrumental settings. Wavelength
calibration was performed via exposures of a thorium-argon lamp 
obtained at the beginning and end of the period of visibility
of the comet. From these arc-lamp exposures, the instrumental
resolution in the order containing the NaD lines was
$0.11$\AA\ FWHM. Due to the short length of the slit in this
observing mode necessary to avoid order overlap ($3.2$ arcsec), no
flux calibration was attempted.\\
\noindent
Because of severe instrumental problems, comparison lamp exposures were
not possible on the 23rd April. Therefore after bias subtraction
and cosmic-ray removal, wavelength calibration was performed using the
strong NaD airglow lines observed simultaneously with the cometary
emission. Exposures of the lunar surface had the reflected solar spectrum
removed, thereby leaving simply the transmission profile of the order
isolating filter. This was then used to correct the
data for the filter transmission profile at all points along the slit.
Flux calibration for these data was obtained via exposures of the standard star
BD $+75^{\circ}325$ (\markcite{s14}Oke 1990) with a $7.0$ arcsec wide slit. A
representative spectrum is shown in figure 2.\\ 

\section{Analysis and Discussion}

\noindent
From these spectra, the brightness and velocity as a function of
distance along the tail
has been modelled assuming that the gas was produced by the nucleus
and/or a near--nucleus source with dimensions much smaller than the
tail length. It then undergoes acceleration in the antisun direction
due to simple radiation pressure, through photon scattering (fluorescence)
in the resonant D-line transitions. In this case the column density of
Na atoms at a sky--projected cometocentric distance $L$ along the tail
is equal to

\begin{equation}
{N({\rm Na})} = {{{Q({\rm Na})}\over{l v(L)}}{{\rm e}^{-{t\over{(r_h^2 \tau)}}}}}\;
\end{equation}

\noindent 
where $Q({\rm Na})$ is the production rate, $\tau$ is the lifetime of the sodium
atoms at a heliocentric distance of 1 AU, r$_h$ is the heliocentric distance
along the tail, $l$ is the sky--projected linear dimension of the tail
cross--section, $v(L)$ is the sky--projected Na velocity along the tail, and
$t$ is the mean travel time of the Na atoms from the nucleus to the tail
distance $L$. In the approximation of a purely antisunward sodium velocity,
then the sky--projected sodium velocity $v(L)$, the heliocentric sodium velocity
$v_h$ and the radial sodium velocity measured in the Echelle spectra $v_E$ are
related by $v(L) = v_h \sin \alpha$ and $v_E = {\dot \Delta} + v_h \cos \alpha$
where $\alpha$ is the phase angle and ${\dot \Delta}$ is the comet geocentric
velocity. If $v_0$ is the starting velocity of the Na atoms from
the source, then the width of the tail will be $l = 2 v_o t + l_o$, where $l_o$
is the linear dimension of the source region. Moreover, the transverse
acceleration of the Na atoms due to the random-walk nature of the scattering
process should be taken into account. We find that Na atoms starting with 
an outflow 
velocity of $v_o \approx 1$ km s$^{-1}$ provides a width $l$ close to that
observed, and that the term $2 v_0 t$ dominates the others for $L > 10^7$ km. 
Then, the average emitted intensity (in rayleighs) of the tail axis is
given by

\begin{equation}
{I({\rm Na})} = {{10^{-6} N({\rm Na}) {{g(L)}\over{r_h^2}}}} = 
{{{{10^{-6}Q({\rm Na})}\over{2 v_o t v(L)}} {{g(L)}\over{r_h^2}}}{\rm e}^{-{t\over{(r_h^2 \tau)}}}}\;
\end{equation}

\noindent 
where $g(L)$ is the photon scattering efficiency, or g--factor, in the D$_1$
and D$_2$ lines at 1 AU.  As this region of the solar spectrum also displays
several additional absorption lines, the g--factor is highly dependent on the 
heliocentric velocity and consequently on  $v(L)$ and $L$. Therefore
the acceleration due to solar radiation pressure, assuming that emission
out the D$_1$ and D$_2$ lines is negligible, is also dependent on the
velocity and heliocentric distance, and allows us to compute the ratio 
between the radiation pressure force and the solar gravitational force;

\begin{equation}
{\beta({\rm Na})} = {{h ~g(L) (1 ~AU)^2}\over{\lambda G M_\odot m}}
\end{equation}

\noindent
where $G$ is the gravitational constant, $h$ is Planck's constant,
$M_\odot$ is the Sun's mass and $m$ is the mass of the Na atom. \\ 
\noindent
Figure 3 shows the g(L) factor, the $\beta(Na)$ factor, 
the modelled and observed tailward velocities and surface brightness
along the tail. Surface brightnesses were measured from the spectra
rather than the CoCAM data due to large uncertainties resulting from
attempts to accurately calibrate these extremely wide-field images.
The lifetime for photoionization of the sodium atom is rather controversial. 
Using measurements of the Na cross section by \markcite{s7}Hudson and Carter 
(1967), \markcite{s8} Huebner et al. (1992) evaluated an experimental 
lifetime of $\tau = 6.17\times10^4$ s at a heliocentric distance of 1 AU. 
However, following a theoretical recalculation of the cross section by 
\markcite{chang} Chang and Kelly (1975), Huebner et al. also suggested a 
value almost three times longer of $\tau = 1.69\times10^5$ s. In previous 
studies of the Moon, Mercury, Io and other comets, values close to the 
shorter "experimental" Na lifetime quoted above have been used.
However, \markcite{s11} Killen et al. (1997) have claimed to require
a longer  lifetime to explain the different accommodation time
of the sodium atoms on  the surface of Mercury and some characteristics
of its atmosphere. Therefore we attempted to fit the observed
brightness distribution assuming both of the above photoionization
timescales. As shown in figure 3, the observed brightness of the tail
is consistent with our model if we adopt the longer
photoionization timescale of $\tau = 1.69\times 10^5~s$ at 1 AU.\\
\noindent
Modelling has also been applied to the CoCAM images, again
assuming that radiation pressure due to resonance fluorescence in
the D--lines is solely responsible for the existence of the tail.
If the particles with the same $\beta$ are
ejected with velocity $v_0$ from the inner coma, then the resulting Na tail
will be a syndynamic tube of width $2 v_0 t$. The axis of this syndynamic tube
is the {\it syndyne} defined by $\beta(Na)$, which is computed by means of
Keplerian mechanics (\markcite{s3}Finson and Probstein 1968,  
\markcite{s5}Fulle 1992). Applying the syndynamic model to CoCAM images
taken on April 21.9 UT and trying to fit the data with different values of
$\beta$, Table 1 lists the Position Angles (PA) of the Na tail:
$PA_O$ is measured from the CoCAM images, and $PA_C$ is the computed
PA of the range of best fitting syndynes of $\beta = 100 \pm 20$. This 
high value of $\beta$ implies that only atoms can satisfy the fit and not 
molecules or dust grains, for which the largest value reported in 
literature is $\beta\simeq2$.\\
\noindent
We applied the same syndynamic model to the D--line emission velocities 
measured within the sodium tail. In Table 2, those observed ($v_{ro}$) on
April 23.9 UT are compared with those predicted from the model ($v_{rc}$),
assuming the best fitting $\beta = 82 \pm 3$. This value is consistent with,
but much more accurate than, that provided by the tail PA fit. We stress that
such a  high $\beta$ value would be impossible for dust particles,
but is entirely reasonable for atoms. Using this value in equation (3) we
deduce ${g/m} = 0.72 \pm 0.03$ s$^{-1}$ amu$^{-1}$. Since for sodium at 1 AU
$g \leq 15.6$ s$^{-1}$, we calculate the mass of any possible
sodium-bearing molecule in the tail as $m \leq 22 \pm 1$ amu. This
conclusively demonstrates that the observed tail is composed of neutral Na
atoms alone, as Na--bearing molecules would then be too massive to be
consistent with such a high $\beta$ ratio. The excellent fit between theory
and observations, provided by both models, allows us to conclude with large
confidence that the observed Na tail is due to fluorescence of
Na atoms which have been released from the nucleus and/or a near--nucleus
source and not {\it in--situ} in the sodium tail. \\
\noindent
The data presented here do not reveal the exact nature of the source
of the sodium atoms observed in the tail.
From the absolute flux calibration of the spectra on April 23.9 UT, we
obtain a production rate of $Q({\rm Na}) \simeq 5\times 10^{25}$ atoms
s$^{-1}$. The water sublimation rate at this time was of
order $\sim 1\times10^{31}$ molecules s$^{-1}$ (\markcite{h2o}Dello
Russo et al. 1997). Assuming that
Hale-Bopp possessed normal cosmic abundances with
Na/O = $2.4\times10^{-3}$ (\markcite{s1} Anders and Grevesse 1989),
then $\sim 0.1$\% of the sodium
in the comet was being released into the tail at this time. Furthermore the
percentage of visible sodium could be altered by a factor 2 or more if we
take into account the very large dust-to-gas ratios mentionned for
Hale-Bopp and the likely significant O content in the refractory dust. In
any case this implies that whatever the source, the majority of Na remained
unobservable in this comet.

\section{Conclusions}

\noindent
We have reported both imaging and high-resolution spectroscopy
of the sodium tail associated with comet C/1995 O1 Hale-Bopp.
Two different approaches have been used to analyze the new 
sodium tail which allow the identification of the mechanism responsible; 
the simple but effective process of fluorescence. Based on this mechanism, 
we expect the appearance of a sodium tail to vary strongly with observing 
geometry (phase angle) and heliocentric velocity of the comet as the Na 
emissions move with respect to the solar Fraunhofer absorption lines
(the Swings effect), thereby changing their intensity and acceleration.
This is therefore a possible explanation
for the different  Na distributions observed in Hale-Bopp at other dates 
(\markcite{s20}Wilson et al. 1997) and in previous comet observations.\\
\noindent
The associated
modelling supports the theoretical value for the photoionization
lifetime of the sodium atom  of $\tau = 1.69\times 10^5~s$ at 1 AU,
a factor 3 higher than that generally used to study the exospheres of
other Solar System bodies, {\it e.g.} the Moon and Io. The tail sodium
atoms appear to have been released in the near-nucleus region of the
comet, and not {\it in-situ} via the break-up of Na--bearing
molecules, ions or dust particles. Assuming normal cosmic abundances
for the comet, $\sim 0.1$\%  of the expected amount
of sodium released by the comet is directly observable
in the Na tail. Both future analysis of existing data and observations
of other comets should concentrate on investigating the exact
source distribution in the inner coma.

\acknowledgments

We gratefully acknowledge the staff of the Isaac Newton Group who
built and operated the  CoCAM. The William Herschel Telescope is 
operated on the island of La Palma by the Royal Greenwich Observatory
in the Spanish Observatorio del Roque de los Muchachos of the
Instituto de Astrofis\`{\i}ca de Canarias. This project has been supported
by the European Commission through the Activity "Access to Large--Scale
Facilities" within the Programme "Training and Mobility of
Researchers", awarded to the Instituto de Astrofmsica de Canarias to
fund European Astronomers' access to its Roque de Los Muchachos and
Teide Observatories (European Northern Observatory), in the Canary Islands.
We further thank M.Mendillo and J.Baumgardner for the use of their
sodium interference filter and their useful suggestions on the observations.\\

\newpage

\begin{deluxetable}{cccc}
\tablecaption{Position angles obtained with the syndynamic model}
\tablehead{
\colhead{L} & \colhead{L}  & \colhead{PA$_O$}  & \colhead{PA$_C$}\\
\colhead{deg}   & \colhead{10$^6$~km} & \colhead{deg} & \colhead{deg}
}
\startdata
1.33  &5.6     &61.9$\pm$1.0 &59.8$\pm$0.5   \nl
2.11  &8.9     &59.2$\pm$0.8 &58.8$\pm$0.5   \nl
2.70 &11.4     &59.0$\pm$0.7 &58.2$\pm$0.5   \nl
3.78 &16.0     &57.1$\pm$0.6 &57.0$\pm$0.5   \nl
4.66 &19.7     &55.6$\pm$0.5 &56.0$\pm$0.5   \nl

\enddata

\tablecomments{
Observed position angles $PA_O$ for the Na tail as a function of the
sky--projected distance $L$ from the nucleus on April 21.9 UT, compared to
those calculated $PA_C$ from the syndynamic model discussed in the text,
with $\beta = 100\pm20$. Note that the model underestimates the observed PA
close to the comet nucleus: to fit these values, larger $\beta$ (i.e. larger
$g$ factors) would be required. However, close to the nucleus the $PA_O$
measurements are affected by the largest errors, due to the Na tail width.}

\end{deluxetable}

\begin{deluxetable}{ccccc}
\tablecaption{Velocities along the sodium tail.}
\tablehead{
\colhead{L} & \colhead{L}  & \colhead{v$_{ro}$} & \colhead{v$_{rc}$} & 
\colhead{t}\\ 
\colhead{deg} & \colhead{10$^6$~km} & \colhead{km~s$^{-1}$} & 
\colhead{km~s$^{-1}$} & \colhead{days} 
}
\startdata
0.89 &3.83     &62$\pm$1     &63$\pm$2      &2.0$\pm$0.1 \nl
3.13 &13.49    &117$\pm$1    &118$\pm$2     &3.7$\pm$0.1 \nl
5.07 &21.89    &149$\pm$1    &151$\pm$3     &5.0$\pm$0.1 \nl
7.17 &31.03    &178$\pm$1    &176$\pm$3     &6.1$\pm$0.1 \nl
\enddata

\tablecomments{
Observed and modelled velocities in the sodium tail on April 23.9 UT.
$L$ is the sky--projected distance from the nucleus, $v_{ro}$ is the measured
radial velocity corrected for the relative velocity between comet nucleus and
Earth, $v_{rc}$ is the computed radial velocity of the Na atoms along
the line of sight in the comet nucleus reference frame (to be compared directly
with $v_ro$) for the best fitting syndyne of $\beta = 82 \pm 3$, $t$ is the
travel time of the Na atoms from nucleus to tail distance $L$.}

\end{deluxetable}

\newpage

\figcaption
{Images of Comet Hale--Bopp obtained with CoCAM and the
narrow--band Na and H$_2$O$^+$ filters on April 16.9 UT. The image is
composed by two square picture with each linear dimension spanning $7.4$
degrees, or $3.1\times10^7$ km at the comet nucleus. The neutral sodium
tail described in the text is seen as a linear feature to the left of the
left picture and distinct from the normal and more diffuse dust tail on the
right. The continuum image has not been subtracted in either
image in order to show clearly that all three tails are distinct
from one another.\label{fig.1}}

\figcaption
{Spectrum of the sodium tail $3.1$ degrees from
the nucleus of Hale-Bopp, obtained with the WHT plus Utrecht
Echelle Spectrograph on April 23.9 UT. Sky lines due to OH and Na
are labelled. The cometary Na D--lines are clearly observable due
to an apparent Doppler shift of $144$ km s$^{-1}$ from the terrestrial
sky emission. \label{fig.2}}

\figcaption
{(a) Theoretical $g$ factors for the D$_1$ line (bottom),
D$_2$ line (middle), and both combined (top)as a function of
the tail distance deprojected along the antisolar direction.
(b) $\beta({\rm Na})$
values in function of the Tail Distance deprojected along the antisolar 
direction. (c) Measured velocities (*), deprojected along the antisolar
direction, compared to the computed velocity (curve). (d) Observed
brightness distribution in the tail as measured from the
high-resolution spectra versus model predictions for two
photoionization lifetimes: 1) (lower) assuming a Na lifetime at 1 AU of
$\tau=14$ hours (the experimental photoionization cross
section)  and 2) (upper) assuming $\tau = 47$ hours
(theoretical photoionization cross section). \label{fig.3}}

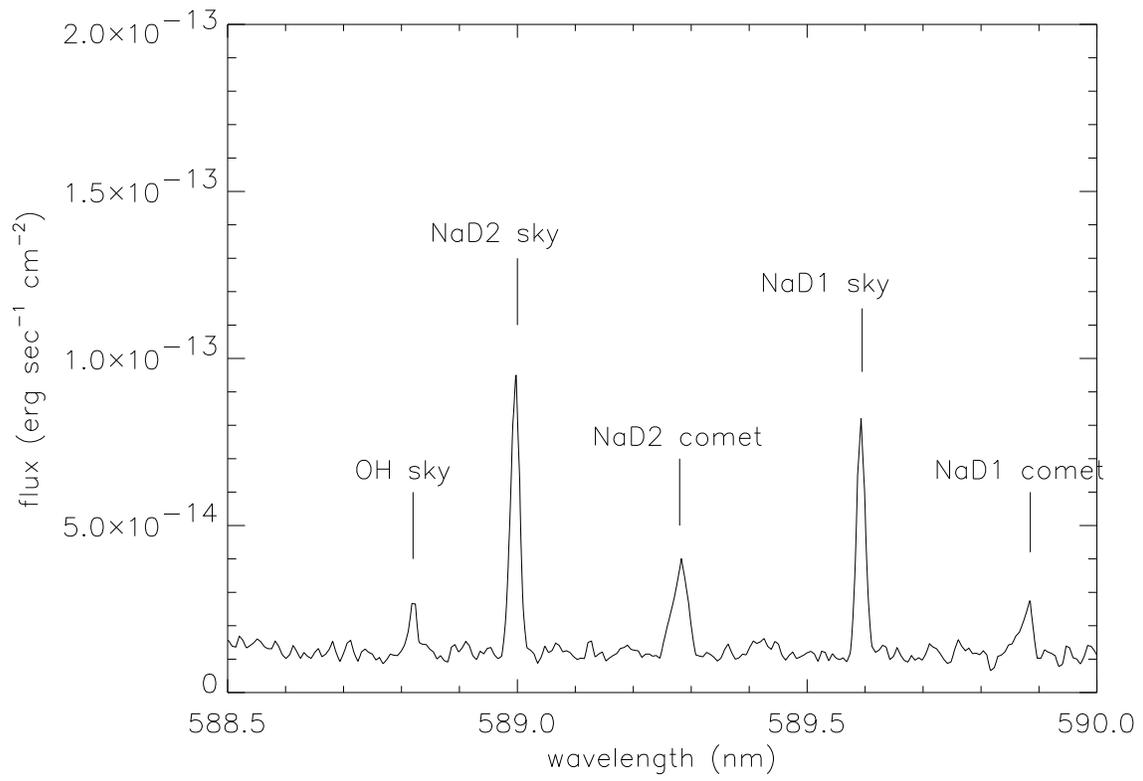
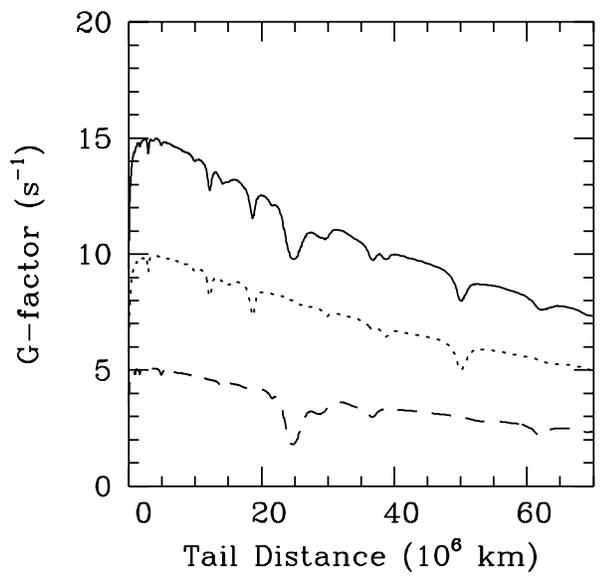
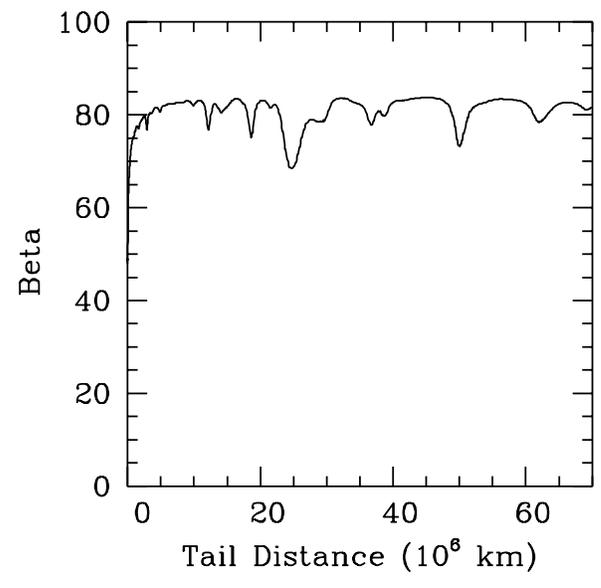
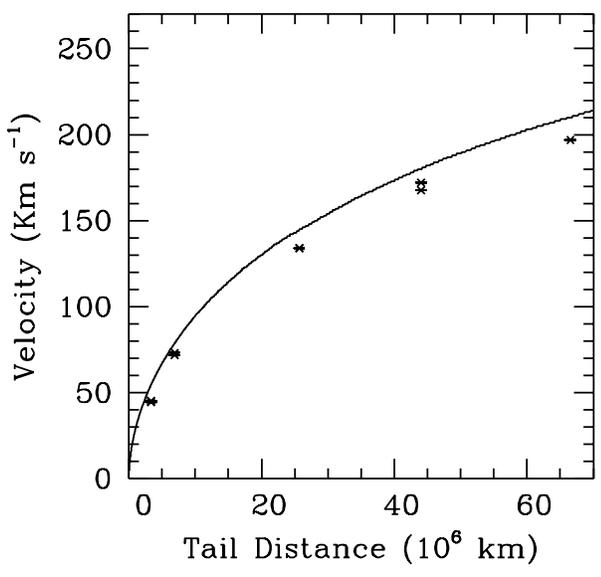
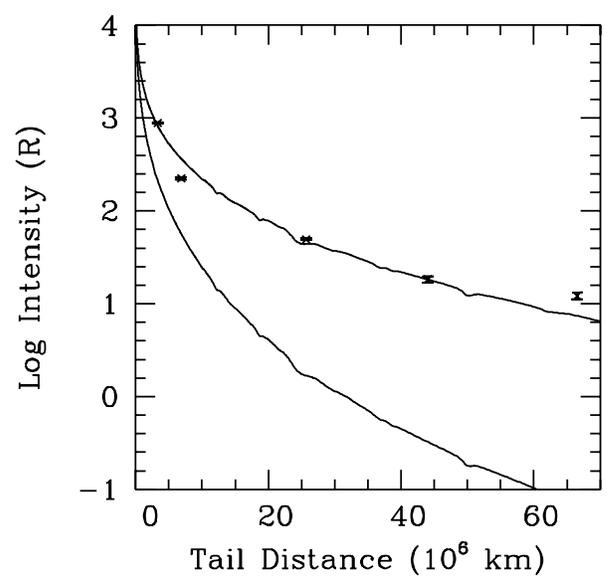
\end{document}